\documentclass[12pt, prd, showpacs]{revtex4}
\usepackage{amsmath}
\usepackage{amssymb}

\begin{document}

\title{Black hole mass formula in the membrane paradigm}
\author{Jos\'{e} P. S. Lemos}
\affiliation{Centro de de Astrof\'isica e Gravita\c c\~ao - CENTRA,
Departamento de F\'{\i}sica, Instituto Superior T\'ecnico - IST,
Universidade de Lisboa - UL, Avenida Rovisco Pais 1, 1049-001,
Portugal, and \\
Gravitational Physics Group,
Faculty of Physics, University of Vienna, Boltzmanngasse 5,
A1090, Wien, Austria}
\email{joselemos@ist.utl.pt}
\author{Oleg B. Zaslavskii}
\affiliation{Department of Physics and Technology, Kharkov
V. N. Karazin National
University, 4 Svoboda Square, Kharkov 61022, Ukraine, and Institute of
Mathematics and Mechanics, Kazan Federal University, 18 Kremlyovskaya St.,
Kazan 420008, Russia}
\email{zaslav@ukr.net}

\begin{abstract}

The membrane paradigm approach adopts a timelike surface, stretched out
off the null event horizon, to study several important black hole
properties.  We use this powerful tool to give a direct derivation of
the black hole mass formula in the static and stationary cases without
and with electric field. Since here the membrane is a self-gravitating
material system we go beyond the usual applicability
on test particles and test fields of the
paradigm.

\end{abstract}

\maketitle



\newpage

\section{Introduction}

The event horizon of a black hole acts like a membrane with
well-defined matter fluid properties.  This was understood first by
Hawking and Hartle from matter fields entering the black hole and
changing its area in a prescribed way \cite{hh72} and Hanni and
Ruffini \cite{hr733} in an analysis of the lines of forces of an
electrically charged particle around a black hole.  It was fully
recognized as a membrane with electric resistivity by Znajek
\cite{znaj} and Damour \cite{dam78} who also showed that the membrane
obeys a Navier-Stokes type equation with surface viscosity
\cite{dam82}. The membrane, being a substitute for the event horizon,
is a lightlike hypersurface in these works.  For practical reasons, it
can be useful to strech slightly the event horizon, into a stretched
horizon, and turn the lightlike membrane into a timelike membrane.
More specifically, the rationale is to replace the event horizon by a
surface in its outside vicinity as if the horizon were stretched, with
its interior being essentially a vacuum spacetime.  This stretched
horizon acts then like a 2-dimensional membrane evolving in time, and
by imposing correct boundary conditions on this membrane, one finds the
desired results.  The stretched horizon is thus a timelike boundary,
which makes this setting prone to using a 3+1 formalism, with three
spatial dimensions and one time dimension. In physical terms, some
processes can be better understood and more intuitive in this timelike
membrane than the lightlike boundary of the event horizon.

This idea of studying the event horizon as a stretched horizon,
through a membrane in a 3+1 spacetime formalism, was devised by Thorne
and collaborators \cite{mt82,zurek85,pt86,spr88} and is called the
membrane paradigm \cite{thorbook86}.  In this 3+1 formalism, a black
hole resembles a star, since now it is realized as a 3-dimensional
space, with an interior, a surface membrane with appropriate boundary
conditions, and an exterior, evolving in time.  Following the approach,
one performs all the calculations in the timelike membrane, located
infinitesimally close to the true black hole horizon, imposes correct
boundary conditions on it, and then finally takes the limit to the
horizon, a lightlike surface, to find the correct desired results.  The
3+1 membrane paradigm is able to recover all the properties
previously found, see \cite{thorbook86}, notably the interpretation of
the membrane as a matter fluid with dissipative properties.  The
membrane paradigm approach has been developed and applied in several
directions. We mention a few.  The black hole complementarity idea was
developed using the membrane paradigm \cite{sussthorla93}, and a
relation to the fuzzball model has been put forward \cite{mathur},
showing that the approach has found echo not only in astrophysics but
also in fundamental physics.  An action for the membrane was devised
and applied to distinct black hole settings \cite{pw1998}, and the
black hole entropy formula was found through an Euclidean action
membrane approach \cite{entlemoszaslavskii2011}. The quasinormal
spectrum of black holes was explored using the membrane paradigm in
\cite{starin2008}, the physics of jets through the membrane
paradigm was studied in \cite{penna2015}, and the paradigm
was applied to a setting with cosmological horizons 
\cite{twang1}.  The Einstein field
equations and the Navier-Stokes equations connection was
developed in \cite{elingf02009}, and it was then realized that the
fluid/gravity duality is in fact a parent of the membrane paradigm
\cite{hubeny2011,bredbergs2012,pft2015,fischkundu2016}. In this
duality, fluids and gravity are similar, like in the membrane paradigm
where the horizon, a pure gravitational entity, behaves as a membrane
composed of matter with fluid behavior, and Einstein equations
can be put in a Navier-Stokes form.  The membrane paradigm was applied
to other theories that contain black holes, namely, to $f(R)$ gravities
\cite{fofrgrav2010}, to the Gauss-Bonnet theory
\cite{jacobsarkar2011}, to the Lovelock theory
\cite{kolekarkothawala2012}, and to Chern-Simons gravity \cite{twang2}.

Now, one interesting and important feature of black holes is that they
possess a simple mass formula.  For the Kerr-Newman family of black
holes, this formula is the Smarr formula \cite{sm}, and it expresses the
mass of a black hole as a bilinear form in terms of the products of
the surface gravity and the area,
the electric potential and the electric charge, and 
the angular velocity and the angular
momentum of the
black hole.  The
formula was generalized to include black holes with surrounding matter
\cite{bch} (see also \cite{cart73,bard73}), and
in an explicit form for extremal black
holes \cite{meinel1}, 
and in addition it was shown by Bardeen, Carter, and Hawking 
that
the Smarr mass formula is connected with the differential
formula for the same quantities as was displayed in the first law of
black hole mechanics \cite{bch}.
The Smarr formula has also been deduced for a
number of different types of black holes, see e.g.  \cite{llp}. 
The black hole
mass formula as found in \cite{bch} is calculated from first principles using
the Komar mass \cite{komar}, a mass definition suited for vacuum
spacetimes with a timelike Killing vector.

Since the membrane paradigm has shown to be a powerful tool for
considering properties in the vicinity of a black hole, it is of
interest to use the membrane paradigm to find from direct principles
the black hole mass formula.  Moreover, in using the membrane
paradigm in this situation, we are paying attention to the properties
of the membrane itself that are connected to its self-gravitation and
thus going beyond the use of the paradigm for test fields in the
vicinity of the horizon of a given black hole background.
Now, to have a
mass formula, one has to start with a mass definition, and for
pure vacuum, one uses the Komar mass \cite{komar}.
In the presence 
of nonvacuum spacetimes with matter field content,
a more suitable mass definition is the Tolman mass
\cite{tolman} (see
also \cite{abreuvisser}),
which further requires that the spacetime is
static or stationary.
Using the Tolman mass definition, 
formulas for black holes were deduced in the quasiblack hole
approach \cite{lemoszaslavskiimass1,lemoszaslavskiimass2}.
To calculate the black
hole mass formula through the membrane paradigm, as is our intention,
the Tolman mass is also the convenient definition.
Given an interior and
an exterior spacetime, the membrane, a matter field, 
reveals itself through the junction conditions \cite{isr,taub,mtw}.  The
membrane is thus a cut between the inner and outer spacetimes that
provides the mass in a consistent manner.  A particularly important
case is when the membrane is at an infinitesimal distance from the
horizon. In such a configuration, one should get the black hole mass
formula.

The aim of the present paper is to give a direct derivation of the
black hole mass formula through the membrane paradigm.  When the
membrane is far from its own gravitational radius, one gets a mass
formula for the membrane spacetime in general.  When the membrane
comes close to its own gravitational radius, i.e., in the horizon
limit, and upon using appropriate boundary conditions which signal the
presence of a horizon, then the membrane formalism yields the black
hole mass formula.  For the inner region, one must note that in the
horizon limit all compatible interior spacetimes, i.e., interior spacetimes
with
appropriate boundary and regularity conditions, give the same mass
formula, so one can
choose the simplest inner spacetime, i.e., Minkowski spacetime.  The
outer spacetime we consider is quite general. We only impose that it
is static or stationary and the spatial sections are topological
spheres.  In this approach it is the Einstein field equation at the membrane
thin layer that conspires to give the black hole mass formula.
In deriving the black hole mass formula,
we consider cases in which there is matter outside the membrane and
the black holes are distorted by outer force perturbations.  We will
use results given in \cite{vis,visprd,tz}.

The paper is organized as follows. In Sec.~\ref{1}, we derive the black
hole mass formula from the Tolman mass for the simplest case, a static
membrane with no electric field.  We do not impose other symmetries,
only staticity.  In Sec.~\ref{2}, we derive the black hole mass formula
from the Tolman mass for an electric shell. Again, we do not impose
other symmetries, only staticity. In the horizon limit, we show that
the electric potential is constant at the horizon.  In Sec.~\ref{3}, we
derive the black hole mass formula from the Tolman mass for a rotating
membrane. We only impose stationarity.  In Sec.~\ref{conc}, we
conclude.

\section{Black hole mass formula in the membrane paradigm:
Static membrane with no electric field}
\label{1}

\subsection{Preliminaries}

\subsubsection{Gravitational field}

Let us consider a 4-dimensional spacetime with
coordinates $x^\mu$ and 
interval $ds$ given by 
\begin{equation}
ds^{2}=g_{\mu\nu}dx^{\mu}dx^{\nu}\,,
\label{mgeneral4}
\end{equation}
where $g_{\mu\nu}$ is the metric
representing the gravitational field
and with $\mu,\nu$ being
spacetime indices, i.e., $\mu,\nu=0,1,2,3$, $0$ being a time
index and $1,2,3$ being spatial indices.
Consider further that the spacetime is static and write the metric
in a 3+1 split as 
\begin{equation}
ds^{2}=-N^{2}dt^{2}+g_{ik}dx^{i}dx^{k}\,,
\label{mstatic0}
\end{equation}
where $N$ is a function of the spatial coordinates
and $i,k=1,2,3$  are the spatial indices.

Let us further consider that the 4-dimensional spacetime contains
an infinitesimally
thin membrane.  The membrane is not necessarily a
2-sphere, it can be some surface with the topology of a
2-sphere.  Inside the membrane,
there is vacuum, and we assume
spacetime there is flat.
The assumption that spacetime inside the membrane is flat
simplifies the calculations, other assumptions
can be provided, giving quantitatively different
results. However, when the membrane is at its own horizon
radius, it does not matter what we have assumed
for the spacetime inside, any compatible inner spacetime
in this limit yields the same results. This is
the basis of the membrane paradigm, namely, one
excises the spacetime interior to the horizon, so it can be any,
and works out the properties of the horizon seen
as a membrane. So, assuming an inner flat spacetime
is an assumption without loss of generality.
Outside the membrane, spacetime is static and consistent with
the existence of a membrane. Let us now state precisely the
complete gravitational field, i.e., the complete spacetime.

Inside the membrane, the vacuum flat spacetime metric can be written
in Gaussian coordinates $(t,l,x^2, x^3)$, from Eq.~(\ref{mstatic0}), 
as
\begin{equation}
ds^{2}=-N_{0}^2dt^{2}+dl^{2}+\gamma_{ab}dx^{a}dx^{b}\,  \label{m0}
\end{equation}
where $N_0$ is some constant, the metric coefficients $\gamma_{ab}$
do not depend on $t$, and
$a,b=2,3$.

Outside, the membrane generates
a generic static metric. In Gaussian coordinates $(t,l,x^2,x^3)$,
the metric of Eq.~(\ref{mstatic0}) takes the form
\begin{equation}
ds^{2}=-N^{2}dt^{2}+dl^{2}+\gamma_{ab}dx^{a}dx^{b}\text{,}  \label{m}
\end{equation}
where the metric coefficients  $N$ and $\gamma_{ab}$
do not depend on $t$ and
$a,b=2,3$. Note that the determinants of the metrics
have the relations $\sqrt{-g}=N\sqrt{g_3}=N
\sqrt{\gamma}$, where $g$
is the
determinant of the spacetime metric (see Eq.~(\ref{mgeneral4})),
$g_3$ is the
determinant of the spatial 3-metric (see Eq.~(\ref{mstatic0})), and 
$\gamma$
is the
determinant of the spatial 2-metric (see Eqs.~(\ref{m0})
and (\ref{m})).

The membrane is located at some $l=l_{0}$ by assumption, and let the
value of $N$ be constant on the membrane, $N=N_0$. This
assumption is not essential but simplifies some formulas.
The matching conditions from the inside to the outside
require, first, that $N_-=N_+=N_0$
on both sides of the membrane, where $-$ corresponds to
the inner side and $+$ corresponds to the outer side at $l_0$,
and, second, that
$\gamma_{ab}^{-}=\gamma
_{ab}^{+}\equiv\gamma
_{0\,ab}$ also on both sides of the membrane.
The matching conditions also require that
the membrane has some stress-energy tensor.

We assume that a priori there is no horizon. When we take
the limit to the membrane gravitational radius, i.e.,
to its own horizon, we obtain a stretched horizon
where the properties of a black hole
can be worked out.

\subsubsection{Energy-momentum tensor}

The spacetime has some  energy-momentum tensor $T_{\mu\nu}$.
We can divide the energy-momentum tensor as
the sum of the inner energy-momentum tensor $T_{{\rm in}\,\mu\nu}$,
the membrane energy-momentum tensor $T_{{\rm membrane}\,\mu\nu}$,
and the outer energy-momentum tensor
$T_{{\rm out}\,\mu\nu}$,
\begin{equation}
T_{\mu\nu}=T_{{\rm in}\,\mu\nu}+T_{{\rm membrane}\,\mu\nu}+
T_{{\rm out}\,\mu\nu}\,.
\label{energmomenttotal}
\end{equation}
Since inside it is Minkowski, the energy-momentum tensor
inside is zero,
\begin{equation}
T_{{\rm in}\,\mu\nu}=0\,.
\label{energmomentin1}
\end{equation}
Since the membrane is infinitesimal, situated at $l_0$,
we cane write the membrane energy-momentum tensor as 
\begin{equation}
T_{{\rm membrane}\,\mu\nu}=S_{\mu\nu}\delta(l-l_0)\,,
\label{tmunumembrane1}
\end{equation}
where $\delta(l-l_0)$ is the Dirac delta function and $S_{\mu\nu}$
is a surface energy-momemtum tensor defined at the membrane.
The outside energy-momentum tensor, with support on
the outer region, can be divided if we wish
into matter and other fields, 
so we can write 
\begin{equation}
T_{{\rm out}\,\mu\nu}=T_{{\rm out\,\,matter}\,\mu\nu}+
T_{{\rm \,\,other\,fields}\,\mu\nu}\,.
\label{energmomentouter1}
\end{equation}
The term $T_{{\rm out\,\,other\,fields}\,\mu\nu}$ can
contain an electromagnetic field,
for instance. In writing Eq.~(\ref{energmomentouter1}), we have
assumed that there are no interaction cross terms between gravity and
matter or between gravity and other fields. This means we assume
minimal coupling throughout.

\subsection{Tolman mass formula for a static
membrane with no electric field}

There are several mass formulas for spacetimes, notably the
Komar mass formula
\cite{komar}
and the
Tolman mass
formula  \cite{tolman}, both 
for static and stationary spacetimes.
The two formulas are related \cite{abreuvisser}
as our calculation will also show. The Tolman mass
formula applies neatly for spacetimes that contain matter.
Since our spacetime possesses a membrane, and so matter,
we use the
Tolman mass formula.
We will first write a mass formula valid for a membrane at
any radius $l_0$ larger than its own gravitational radius.
Then, afterward, we will apply this formula
to the case when the
membrane is at its own gravitational radius, i.e., at the horizon.
We will then find the black hole mass formula.

The Tolman mass for a given static spacetime
with a stress-energy tensor $T_{\mu\nu}$ 
is defined as 
\begin{equation}
M=\int_\Sigma(-T_{0}^{0}+T_{k}^{k})\sqrt{-g}\,d^{3}x
\label{mass4}
\end{equation}
with $g$ being the determinant of the metric $g_{\mu\nu}$
in Eq.~(\ref{mgeneral4}),
$d^3x=dx^1dx^2dx^3$,
and the integral is performed
over a 3-space $\Sigma$ with $t={\rm constant}$.
In the metric Eq.~(\ref{mstatic0}),
the mass formula is 
$
M=\int_\Sigma(-T_{0}^{0}+T_{k}^{k})\,N\,\sqrt{g_{3}}\,d^{3}x
$,
with $g_{3}$ being the determinant of the metric $g_{ik}$.
In addition, since $g_3=\gamma$ where 
$\gamma$ is the determinant of the 2-metric $\gamma_{ab}$
we have
\begin{equation}
M=\int_\Sigma(-T_{0}^{0}+T_{k}^{k})\,N\,\sqrt{\gamma}\,d^{3}x\,.
\label{massgamma}
\end{equation}
We will resort to these formulas, mainly to
Eqs.~(\ref{mass4}) and (\ref{massgamma}), for finding the Tolman mass.

Now, there are clearly tree distinct
regions, as we have made explicit in the splitting of the energy-momentum
tensor, Eq.~(\ref{energmomenttotal}). So, here, one should
also 
split the mass formula into three parts, namely, the inner mass $M_{\rm in}$,
the membrane mass $M_{\rm membrane}$, and the outside mass $M_{\rm out}$. Thus, we write
\begin{equation}
M=M_{\rm in}+M_{\rm membrane}+M_{\rm out}\,,
\label{sum0}
\end{equation}
and calculate the contribution of each part to the total
spacetime mass $M$.

Inside is vacuum, so from Eqs.~(\ref{energmomentin1}) and (\ref{massgamma})
one has 
\begin{equation}
M_{\rm in}=0\,.
\label{in}
\end{equation}

The membrane term $M_{\rm membrane}$
is quite interesting.
From Eq.~(\ref{massgamma}) we write the
contribution to the membrane mass as
\begin{equation}
M_{{\rm membrane}}=\int_{{\rm membrane}}\left(-{T_{\rm membrane\, 0}}^{\,\,0}+
{T_{\rm membrane\, k}}^{\,\,k}
\right)
\,N\sqrt{\gamma}\,d^{3}x\,\,.
\label{x1}
\end{equation}
Since the membrane is infinitesimally thin,
there is the delta contribution given in Eq.~(\ref{tmunumembrane1}).
Moreover, due to the 2-dimensional character of the
membrane, there are no radial stresses, i.e., 
$S_{l}^{\,\,l}=0$.
Performing then the integral over $l$ across the membrane in
Eq.~(\ref{x1}),
we get
\begin{equation}
M_{{\rm membrane}}=
\int_{{\rm membrane}}
(-S_{0}^{\,\,0}+S_{a}^{\,\,a})\,N\,dS\,\,,
\label{x2}
\end{equation}
where $dS =\sqrt{\gamma(l_0)}\,d^{2}x$, and $d^2x=dx^2dx^3$.
Given an outside spacetime and an inside one, 
one has from the junction conditions \cite{isr,mtw,taub} the relation $8\pi
S_{\mu}^{\nu}=[[K_{\mu}^{\nu}]]-\delta_{\mu}^{\nu
}[[K]]$, where $K_{\mu}^{\nu}$ is the extrinsic curvature
tensor, and the symbol $[[...]]$ means  $[[...]]=[(...)_{+}-(...)_{-}]$
with $+$ being the value of the relevant quantity at the
membrane from the outside and $-$ being the value of the same relevant
quantity at the
membrane from the inside. 
In our case, we find  $8\pi
\left(-S_{0}^{0}+S_{a}^{a}\right)=-2[[K_{0}^{0}]]$. Put $n^{\mu}$ as the unit
vector normal to the membrane.
Since $K_{\mu\nu}=-\nabla_\nu n_\mu$, where
$\nabla_\mu$ denotes covariant derivative,
and $n_\mu=\frac{\partial l}{\partial x^\mu}$ we can calculate
$K_{0}^{0}$.
Indeed, $K_{0}^{0}=-\nabla_0{n^{0}}
=- \frac{1}{N}\frac{\partial N}{\partial l}$. As a result, we
obtain
$8\pi
(-S_{0}^{0}+S_{a}^{a})=\frac{2}{N}\left[ 
\left( \frac{\partial N}{\partial l}\right)
_{+}-\left( \frac{\partial N}{\partial l}\right)_{-}\right]$,
but since inside is Minkowski and $N=N_0$, the expression becomes
$8\pi
(-S_{0}^{0}+S_{a}^{a})=\frac{2}{N}
\left( \frac{\partial N}{\partial l}\right)_{+}$. Thus, the membrane mass is
\begin{equation}
M_{\rm membrane}=\int_{\rm membrane}
\sigma \,dS \,,
\label{al}
\end{equation}
where 
\begin{equation}
\sigma=
\frac{1}{4\pi}
\left(\frac{\partial N}{\partial l}\right)_{+}\,
\label{muresult}
\end{equation}
is the surface mass density of the membrane.

The contribution 
from the outside can be written as
\begin{equation}
M_{{\rm out}}=\int_{\rm out}\left(-{T_{\rm out\,0}}^{\,0}+
{T_{\rm out\,k}}^{\,k}\right)
\,N\,\sqrt{\gamma}\,d^{3}x\,,
\label{out}
\end{equation}
where one can further split into matter fields and other fields,
as we have done in Eq.~(\ref{energmomentouter1}).

Thus, putting together
Eqs.~(\ref{in}), (\ref{al})-(\ref{muresult}), and~(\ref{out}), into 
Eq.~(\ref{x1}),
we obtain the mass formula from the Tolman definition for a
spacetime with a membrane.
As an example, if the outside spacetime is vacuum Schwarzschild, then
$M_{\rm in}=0$, $M_{{\rm membrane}}=M$, and 
$M_{\rm out}=0$, so the Tolman formula gives $M=M_{{\rm membrane}}=M$. 
It is
interesting to note that Tolman formula puts all the mass, i.e., membrane
rest mass plus gravitational mass, in the membrane itself.

\subsection{Black hole mass formula in the
membrane paradigm with no electric field}

\subsubsection{Introduction}

The Tolman formula is thus valid for a spacetime with a membrane
and without horizons. We now want to test the Tolman formula in the strongest
gravitational field possible, i.e., in the black hole limit.
In this limit we should take $N_0\rightarrow 0$, where
$N_0\rightarrow 0$ is the value of $N$ at the membrane.

Before we proceed, a remark is in order.  In the above treatment, we
assumed that the spacetime metric inside the membrane is the Minkowski
metric.
However, in the black hole limit, our results do not change if
for the interior region we choose some other metric compatible with the
exterior.
Indeed, the continuity of the metric implies the boundary
condition $N_{-}=N_{+}$ where for simplicity we have chosen
$N_{-}=N_{+}=N_{0} ={\rm constant}$, on the boundary.  The black hole
limit under discussion means $N_{0}\rightarrow 0$. Making a
nonrestrictive assumption that $\frac{\partial N}{\partial l}\geq 0$,
which holds for all reasonable spacetimes under consideration, one has
that everywhere inside $N\rightarrow 0$ uniformly as well. Therefore,
$\left(\frac{\partial N}{\partial l}\right) _{-}\rightarrow 0$, and we
again obtain Eq.~(\ref{al}) together with Eq.~(\ref{muresult}).

\subsubsection{Black hole mass formula in the membrane
paradigm with no electric field}

In the black hole limit, write
$l_{\rm h}\equiv l_{0\,\rm at\, the\, horizon \,limit}$,
and note that indeed $N\to0$.
However, the possible problematic
part for the mass, the membrane part
given by Eqs.~(\ref{al}) with (\ref{muresult}),
does not depend on $N$,
rather, it depends on $\frac{\partial N}{\partial l}$.
So, we have to calculate $\frac{\partial N}{\partial l}$
from the outside at the horizon limit.
For that, we observe that for a regular horizon the metric potential $N$
just outside the horizon must obey \cite{vis}
\begin{equation}
N=\kappa (l-l_{\rm h})+a(x^{a})(l-l_{\rm h})^{3}+...
\label{Nkappal-l01st}
\end{equation}
where $\kappa$ is the surface gravity, a constant,
and
$a(x^{a})$ is a function of the angular coordinates $x^a$.
Thus,
\begin{equation}
\kappa=\lim_{N\to0}\left(\frac{\partial N}{\partial l}\right)_{+}\,.
\label{kappa1}
\end{equation}
Comparing Eqs.~(\ref{muresult})
and (\ref{kappa1}),
we find also that $
\lim_{N\to0}
\sigma= \frac{1}{4\pi}
\kappa
$. So, using 
that
$\kappa$ is a constant, we can
perform the integral in 
Eq.~(\ref{al})
to obtain
\begin{equation}
M_{{\rm membrane\, at \,the\,horizon}}= \frac{1}{4\pi}\kappa A\,,   \label{s}
\end{equation}
where $A\equiv A_{\rm h}$ is the horizon area.

Inserting Eqs.~(\ref{in}), (\ref{out}), and (\ref{s})
into Eq.~(\ref{sum0})
yields the black hole mass formula
\begin{equation}
M=\frac{1}{4\pi}\kappa A+M_{\rm out}\,.
\label{tot}
\end{equation}
This black hole mass formula obtained
through the membrane paradigm approach
is the same as that obtained by other methods 
\cite{bch} (see also \cite{cart73,bard73}).
When $M_{\rm out}=0$, it is the Smarr formula for a static,
i.e.,  Schwarzschild, black hole in general relativity \cite{sm}.

Price and Thorne \cite{pt86}
in their detailed paper on the membrane paradigm,
do not arrive at our Eq. (21) for the mass formula of a black
hole. The key point in our derivation is the use of the Tolman mass
definition which includes under its integral, not only the $T^0_0$
component at the membrane, but also the membrane $T^k_k$ stress
components. These latter contribute decisively to the integral at the
horizon and thus to the black hole mass.

\subsubsection{Interpretation of the surface gravity}

The surface gravity of a body is defined as the acceleration of a test
particle at the body's surface. If the body is a black hole 
this acceleration at the black hole's surface, the horizon,
is infinite, but the normalized surface gravity, or simply
the black hole's surface gravity $\kappa$,
defined as
the proper acceleration times the redshift factor at the horizon,
is finite. 
The black hole's surface gravity $\kappa$
also defines the Hawking
temperature.

But now note that, as a byproduct of the formalism we have developed
within the membrane paradigm, we
have obtained one more
interpretation for the surface gravity. Indeed,
comparing Eqs.~(\ref{muresult})
and (\ref{kappa1})
and using also Eq.~(\ref{s}),
we can write
\begin{equation}
\lim_{N\to0}
\sigma=
\frac{1}{4\pi}
\kappa=
\frac{M_{{\rm membrane\, at \,the\,horizon}}}{A}\,,
\label{inter}
\end{equation}
i.e., the quantity $\kappa$ is nothing other than $4\pi$ times the
surface energy density on the membrane in the horizon limit. This
surface density is a constant in the limit under discussion.  Such an
interpretation of the surface gravity as a surface energy density,
given in Eq.~(\ref{inter}), could not be given in terms of a true
black hole since a true black hole does not have any membrane on the
horizon at all and the mass formula is obtained from a different
approach. Indeed, in the black hole approach, the mass is defined at
infinity and thus cannot be interpreted as localized at the horizon.

Our interpretation relies
on self-gravitating effects since the expression for the membrane's
surface
stress-energy tensor implies the validity of the Einstein equations
\cite{isr,taub,mtw}.
Indeed,
Eq.~(\ref{inter})
is a reflection of the
Einstein equation,
with $\frac{1}{4\pi}
\kappa$ being related to the gravitational, geometrical,
part of the equation and
$\frac{M_{{\rm membrane\, at \,the\,horizon}}}
{A}$
being related to the matter,
energy-momentum tensor, part of the equation.
Equation~(\ref{inter})
with the interpretation of the surface gravity $\kappa$
as a surface energy density $\sigma$ is new. Indeed, such an interpretation
can be put forward only after one finds the black hole mass formula,
Eq.~(\ref{tot}), through the membrane paradigm, as we did.

\section{Black hole mass formula in the membrane paradigm:
Static membrane with electric field}
\label{2}

\subsection{Preliminaries}

\subsubsection{Gravitational and electric fields}

We assume the same type of membrane configuration and
the same type of 
gravitational field as in the previous section, so
that Eqs.~(1)-(4) still hold. But now, we consider
further that there is in addition
an electric field, and so one should consider
the Einstein-Maxwell equations for finding a
mass formula for the spacetime with an electric membrane
and a mass formula in the black hole limit.

Since the gravitational part has been treated
in the last section, we now analyze the electromagnetic field which
has its own special features.
The electromagnetic field is characterized by
a generic antisymmetric Maxwell tensor $F_{\mu \nu}$.
The Maxwell equations for $F_{\mu \nu}$ are
\begin{equation}
\nabla_\nu F^{\mu \nu}=4\pi j^\mu\,,
\label{max00}
\end{equation}
i.e.,
\begin{equation}
\frac{1}{\sqrt{-g}}
\frac{\partial (\sqrt{-g}F^{\mu \nu})}{\partial x^{\nu}}=4\pi
j^{\mu}\,.
\label{maxagain}
\end{equation}
where $\nabla_\nu$ denotes covariant derivative
and $j^{\mu}$
is the generic electric 4-current.
There is another set of Maxwell equations,
namely,
\begin{equation}
\frac{
\partial\,}
{\partial x^\alpha}
\left( \varepsilon^{\alpha\beta\mu\nu}
F_{\mu \nu}\right)=0\,
\label{maxagain2}
\end{equation}
where $\varepsilon^{\alpha\beta\mu\nu}$
is the Levi-Civita tensor.
This last set of equations, Eq.~(\ref{maxagain2}),
in turn, permits the electromagnetic Maxwell field $F_{\mu \nu}$
to be written in terms of a 
4-potential $A_{\mu}$ as 
\begin{equation}
F_{\mu \nu}=\frac{\partial A_{\nu}}{\partial x^{\mu}}-\frac{
\partial A_{\mu}}{\partial x^{\nu}}\,.
\label{fpotdef}
\end{equation}
The 4-potential $A_{\mu}$ can be split as 
$A_{\mu}=(\varphi,A_i)$, for some
electric potential $\varphi$ and a 3-potential $A_{i}$.
For an electric
ansatz, as we will assume, $A_i=0$, and so 
\begin{equation}
A_\mu=\varphi\,\delta^0_\mu\,,
\label{pot0}
\end{equation}
where $\delta^\nu_\mu$ is the Kronecker delta.
Putting Eq.~(\ref{pot0}) into Eq.~(\ref{fpotdef})
one finds that the only nonzero components of $F_{\mu \nu}$
are $F_{0i}$, i.e., 
\begin{equation}
F_{0i}=-\frac{
\partial \varphi}{\partial x^{i}}\,.
\label{electricansmax}
\end{equation}
This confirms that there is pure electric field, since
the components $F_{0i}$ are related to it.
All this formulation is generic for an electric ansatz,
and since the problem is static, the quantities are
time independent.

We now specify that the interior
spacetime is a flat spacetime with
no electric
field, that there is an electric
membrane, and that there is some electric field outside
compatible with the existence of the membrane.
For the inside, we have a zero Maxwell tensor,
and for the outside, we have a Maxwell tensor that
we write as $F_{\mu \nu}^{\rm out}$, having in 
mind that $F_{0i}^{\rm out}$ are the only
nonzero components from Eq.~(\ref{electricansmax}).
We can then write an expression for the
Maxwell tensor field $F_{\mu \nu}$,
valid throughout the whole spacetime, such that
it represents zero Maxwell tensor
inside the membrane, and $F_{\mu \nu}^{\rm out}$
at and outside it.
Thus, in the spirit of the membrane paradigm, 
we write for the Maxwell tensor 
the following expression,
\begin{equation}
F^{\mu\nu}={F}^{\mu \nu}_{\rm out}\theta (l-l_{0})\,,
\label{maxtot}
\end{equation}
where $\theta(l-l_{0})$ is the step function.
Equation~(\ref{maxtot}) has implicitly in it that there are three
regions, the inner, the membrane, and the outer regions. It has also
implicitly in it, due to the $\theta$ function, that inside we have
$F_{\mu \nu}^{\rm in}=0$.  Then, since Eq.~(\ref{maxtot}) is a
product, and the Maxwell equation, Eq.~(\ref {maxagain}), involves a
derivative for $F^{\mu \nu}$, the current $j^{\mu }$ is a sum of two
currents, namely, \begin{equation} j_{\rm
membrane}^{\mu}+{j}^{\mu}_{\rm out}=j^{\mu}\,.  \end{equation} Let us
analyze each current at a time.

For the membrane current, $j_{\rm{membrane}}^{\mu}$, 
from the Maxwell equations, Eq.~(\ref{maxagain}),
and the specific form of the Maxwell field given in Eq.~(\ref{maxtot}),
we have that
\begin{equation}
j_{\rm{membrane}}^{\mu}=\frac{1}{4\pi}
F^{\mu \nu}_{\rm{out}}\delta (l-l_{0})\frac{\partial l}{
\partial x^{\nu}}\,.
\label{curgen}
\end{equation}
This term is very important and can yield a singular
term when one lowers the membrane to its own
gravitational radius or horizon.
Instead of $j_{\rm{membrane}}$, this current could be 
called $j_{\rm{sing}}$
since it can yield a singular term
at the horizon.
Recalling the electric
anzatz, $A_i=0$, and that
the only components of $F_{\mu \nu}$
that do not
vanish are $F_{0i}$, see Eq.~(\ref{electricansmax}),
we find that the only nonzero component in Eq.~(\ref{curgen}) is
\begin{equation}
j_{\rm membrane}^{0}=\frac{1}{4\pi}
F^{0l}_{\rm{out}}\delta (l-l_{0})\,.
\label{j0final}
\end{equation}
Since $F^{0l}$ corresponds to a radial electric field $E^l$,
say, 
we see that at the membrane there is an electric
current related to the radial electric field.
From this equation, we can find
an expression
for the total electric charge $Q$
at the membrane, $l=l_0$. This will take a little detour.
For some generic 4-current $j^\mu$, the electric charge is defined as
$Q=\int_{\Sigma}j^\mu d\Sigma_{\mu}$,
where $d\Sigma_{\mu}
\equiv\frac{1}{3!}
\varepsilon_{\mu\alpha\beta\gamma}
\left[
\frac{\partial(x^\alpha,x^\beta,x^\gamma)}
{\partial(a,b,c)}\right]
dadbdc$
is the 3-dimensional volume element of a
hypersurface $\Sigma$ parametrized by $(a,b,c)$
through $x^\alpha=x^\alpha(a,b,c)$,
$\left[\frac{\partial(x^\alpha,x^\beta,x^\gamma)}
{\partial(a,b,c)}\right]$ is the $3\times3$ Jacobian determinant,
and $\varepsilon_{\mu\alpha\beta\gamma}$ is the
Levi-Civita
totally antisymmetric tensor which can be written
as $\varepsilon_{\mu\alpha\beta\gamma}=\sqrt{-g}\,
\epsilon_{\mu\alpha\beta\gamma}$
with $\epsilon_{\mu\alpha\beta\gamma}$ being the alternating symbol
or Levi-Civita tensor density
(being equal to $1$ if $\mu$, $\nu$, $\alpha$, and $\beta$ is an even permutation
of 0, 1, 2, 3, equal to $-1$ if it is an odd permutation
of 0, 1, 2, 3, and equal to zero otherwise).
Here, since the only nonzero component of $j^\mu$ is 
$j^0$, see Eq.~(\ref{j0final}),
we have $Q=\int_{\Sigma}j^0 d\Sigma_0$. Parametrizing
$\Sigma$ by $a=l$, $b=x^2$, and $c=x^3$, we have 
$Q=\int_{\Sigma} j^0 \sqrt{-g}\,dl\,dx^2\,dx^3$, or using Eq.~(\ref{j0final}),
we get
$Q=\frac{1}{4\pi}\int_{\Sigma} F^{0l}_{\rm{out}}\delta (l-l_{0})
N\sqrt{\gamma}dl\,d^2x$,
where $d^2x=dx^2\,dx^3$.
Integrating through the delta function, it becomes 
$Q=\frac{1}{4\pi}\int_{\rm membrane} F^{0l}_{\rm{out}}(l_0)
N(l_0)\sqrt{\gamma(l_0)}\,d^2x$.
Defining $dS\equiv \sqrt{\gamma(l_0)}\,d^2x$ as the invariant
area element on the 2-dimensional membrane,
we have 
$Q=\frac{1}{4\pi}\int_{\rm membrane} F^{0l}_{\rm{out}}(l_0) N(l_0)dS$.
Interesting to note that this result can also be found
without the use of the $\delta$-function. Indeed, 
from 
$Q=\int_{\Sigma}j^\mu d\Sigma_{\mu}
=\frac{1}{4\pi}\int_{\Sigma} \nabla_\nu{F^{\mu\nu}}d\Sigma_{\mu}$,
where Eq.~(\ref{max00}) has been used,
one can invoke 
Stokes' theorem
which
for this case reads
$2\int_{\Sigma} \nabla_\nu{F^{\mu\nu}}d\Sigma_{\mu}
=\int_S F^{\mu\nu}dS_{\mu\nu}$,
yielding 
$
Q=\frac{1}{8\pi}\int_S F^{\mu\nu}dS_{\mu\nu}
$,
where 
$dS_{\mu\nu}
\equiv \frac12 \varepsilon_{\mu\nu\alpha\beta}
\left[
\frac{\partial(x^\alpha,x^\beta)}
{\partial(e,f)}\right]
dedf$
is the
2-dimensional area element of
the boundary of $\Sigma$, namely, a
2-surface $S$ parametrized by $x^\alpha=x^\alpha(e,f)$, 
with
$\left[\frac{\partial(x^\alpha,x^\beta)}
{\partial(e,f)}\right]$ being the $2\times2$ Jacobian determinant.
Then, one finds again 
$Q=\frac{1}{4\pi}\int_{\rm membrane} F^{0l}_{\rm{out}}(l_0) N(l_0)dS$.
Thus, we can write for the total charge
\begin{equation}
Q=
\int_{\rm membrane} \sigma_{\rm e}\, dS\,,
\label{totalcharge}
\end{equation}
where  $\sigma_{\rm e}$ is the
membrane's electric charge surface density, given by
\begin{equation}
\sigma_{\rm e}=\frac{1}{4\pi}
F^{0l}_{\rm{out}}N\,,
\label{gauss0}
\end{equation}
and where $F^{0l}_{\rm{out}}N$
is evaluated at the membrane $l_0$ from the outside.

For the outside current, $j_{\rm{out}}^{\mu}$,
we have
from the Maxwell equations, Eq.~(\ref{maxagain}),
and the form of the Maxwell field, Eq.~(\ref{maxtot}),
the following equation
$
\frac{1}{\sqrt{-g}}
\frac{\partial (\sqrt{-g}F^{\mu \nu}_{\rm out})}{\partial x^{\nu}}=4\pi
j^{\mu}_{\rm out}
$.
If there are only electric fields this equation yields
\begin{equation}
\frac{1}{\sqrt{-g}}
\frac{\partial (\sqrt{-g}F^{0 k}_{\rm out})}{\partial x^{k}}=4\pi
j^0_{\rm out}\,.  \label{maxagainout}
\end{equation}

\subsubsection{Electromagnetic energy-momentum tensor}

The formulas for the energy-momentum tensor
given in Eqs.~(\ref{energmomenttotal})-(\ref{energmomentouter1})
are still valid when one has a static spacetime with
an electric field. 
However, it is important to isolate now the electric 
 part of the
energy-momentum tensor,
$T_{\quad 
\mu}^{{\rm em}\,\,\nu}$. By definition,
\begin{equation}
T_{\quad \mu}^{{\rm em}\,\,\nu}=\frac{1}{4\pi}(F_{\alpha}^{\; \mu}F_{\nu
}^{\;\alpha}-\frac{1}{4}\delta_{\mu}^{\;\nu}F_{\alpha \beta}F^{\alpha \beta
}),  \label{set}
\end{equation}
where $F_{\mu \nu}$ is the Maxwell tensor. In our case, since
$F_{0i}$ are the only components that do not
vanish, the electromagentic energy-momentum tensor has the
following nontrivial components,
\begin{equation}
T_{\quad 0}^{{\rm em}\,0}=-\frac{1}{8\pi N^{2}}F_{0i}F_{0j}g^{ij}\,,
\label{tem1}
\end{equation}
\begin{equation}
T_{\quad j}^{{\rm em}\,i}=\frac{1}{8\pi N^{2}}(F_0^{\,k}F_{0k}
\delta_{j}^{\;i}-2F_0^{\,i}F_{0j})\,.
\label{tem2}
\end{equation}

Since we are after a mass formula, and we use the Tolman mass definition given
in Eq.~(\ref{massgamma}), we still need
to develop in the mass formula the electromagnetic part of the
energy-momentum tensor.

\subsection{Tolman mass formula for an electric membrane}

To calculate the contribution to the mass from the electric charge
of the membrane, we use the results above.
From Eqs.~(\ref{tem1}) and (\ref{tem2}),
we have that the combination $-T_{\quad 0}^{{\rm em}\,0}+
T_{\quad k}^{{\rm em}\,k}$ that enters into the Tolman
formula, Eq.~(\ref{massgamma}), is
$-T_{\quad 0}^{{\rm em}\,0}+
T_{\quad k}^{{\rm em}\,k}=-\frac{1}{4\pi}
\,F_{0k}\,F^{0k}$.
It follows from Eq.~(\ref{massgamma})  that the
contribution of the electromagnetic field
to the electromagnetic mass $ M_{{\rm em}}$ is
\begin{equation}
M_{{\rm em}}= \frac{1}{4\pi}
\int_\Sigma F_{0k}F^{0k}N\,\sqrt{\gamma}\,d^{3}x\,.
\label{mem1}
\end{equation}
From Eq.~(\ref{electricansmax}), i.e.,
$F_{0k}=-\frac{
\partial \varphi}{\partial x^{k}}$,
we have
$
M_{{\rm em}}= -\frac{1}{4\pi}
\int_\Sigma \varphi_{,k}F^{0k}N\,\sqrt{\gamma}\,d^{3}x
$,
where the integration is over the whole 3-space $\Sigma$,
and to simplify the notation we use here ${}_{,k}\equiv\frac{
\partial\;}{\partial x^{k}}$.
Performing the integral by parts gives 
$-\int_\Sigma  \varphi_{,k}F^{0k}N\,\sqrt{\gamma}\,d^{3}x
=-\int_\Sigma  \left(\varphi F^{0k}N\,\sqrt{\gamma}\right)_{,k}\,d^{3}x
+\int_\Sigma  \varphi \left(F^{0k}N\,\sqrt{\gamma}\right)_{,k}\,d^{3}x
$. From Gauss' theorem the first term can be swapped to
a surface integral at infinity,
$-\int_\Sigma  \left(\varphi F^{0k}N\,\sqrt{\gamma}\right)_{,k}\,d^{3}x
=-\int_{S_\infty}(\varphi
F^{0k}N)n_{k}dS$,
where $n_k$ is the normal to the two surface at infinity, $S_\infty$.
Now, this
surface term  vanishes because at infinity
one has
$N\rightarrow 1$, $F^{0k}\sim \frac{1}{l^{2}}$, $dS \sim l^{2}$, $\varphi \sim 
\frac{1}{l}$.
Thus, $M_{{\rm em}}$ reduces to
$M_{{\rm em}}= \frac{1}{4\pi}
\int_\Sigma \varphi \left(F^{0k}N\,\sqrt{\gamma}\right)_{,k}\,d^{3}x$.
From Eq.~(\ref{maxtot}), we can then write
$M_{{\rm em}}= \frac{1}{4\pi}
\int_\Sigma \varphi \left(F^{0k}_{\rm out}\theta(l-l_0)
N\,\sqrt{\gamma}\right)_{,k}\,d^{3}x$.
So, splitting $\theta(l-l_0)$, we have
$M_{{\rm em}}= \frac{1}{4\pi}
\int_\Sigma  \varphi_{\rm out} F^{0l}_{\rm out}\delta(l-l_0)
N\,\sqrt{\gamma}\,d^{3}x
+\int_{\Sigma_{\rm out}} \varphi_{\rm out} \left(F^{0k}_{\rm out}
N\,\sqrt{\gamma}\right)_{,k}\,d^{3}x
$, where $\varphi_{\rm out}$ is the electric potential in the
outer region and  $\Sigma_{\rm out}$ 
is the outer 3-space.
The first term is related to $j^0_{\rm membrane}$
and after taking care of the $\delta$ function can be written as
$\frac{1}{4\pi}\int_{\rm membrane} \varphi_{\rm out}  F^{0l}_{\rm out}
N\,dS$ with $dS=\sqrt{\gamma}\,d^2x$, and the second term
is related to $j^0_{\rm out}$.
Thus, we can divide $M_{{\rm em}}$ as
\begin{equation}
M_{{\rm em}}=M_{{\rm membrane\,\, em}}+M_{{\rm  out\,\, em}}
\,,
\label{massem}
\end{equation}
with
\begin{equation}
M_{{\rm  membrane\,\, em}}=\int_{\rm membrane} \varphi_{\rm out} \,\sigma_{\rm e}\,dS
\,,
\label{massmembranegeneral}
\end{equation}
where
$\sigma_{\rm e}=\frac{1}{4\pi}
F^{0l}_{\rm{out}}N$ as in Eq.~(\ref{gauss0}),
and 
\begin{equation}
M_{{\rm  out\,\, em}}=\int_{\Sigma_{\rm out}} \varphi_{\rm out} j^0_{\rm out}
N\,\sqrt{\gamma}\,d^3x
\,.
\label{outerem1}
\end{equation}
This is the contribution of the electromagnetic mass to the Tolman mass.

The full mass $M$ is then given by the sum of the 
membrane gravitational matter contribution, Eqs.~(\ref{al})-(\ref{muresult}),
plus the membrane electrical contribution, Eq.~(\ref{massmembranegeneral}),
plus the outer contribution involving all matter and nonelectric fields,
Eq.~(\ref{out}),
plus the outer electric contribution
Eq.~(\ref{outerem1}).

\subsection{Black hole mass formula in the membrane paradigm with electric field}

\subsubsection{Constancy of the electric potential at the horizon}

When we want to treat the black hole limit, i.e., 
take $N\rightarrow 0$, we have to be careful with the physical
quantities so that they do not blow up at the horizon.
So, to avoid problems and confusion caused by pure coordinate
effects when the original frame becomes degenerate in the horizon or
near-horizon
limit, it is convenient to use normalized components in an orthonormal
basis. Thus, we must use normalized variables, the hat variables,
such that $X_{\hat{0}}=\frac{X_{0}}{N}$. For the electric
components $F_{{\rm out}\,0i}$, the only components of
$F_{{\rm out}\,\mu\nu}$ that exist
in this problem, we have 
from Eq.~(\ref{electricansmax}) that
$
F_{{\rm out}\,\hat{0}i}=\frac{F_{{\rm out}\,0i}}{N}=-\frac{1}{N}
\frac{
\partial \varphi_{\rm out}}{\partial x^i}
$,
and these quantities must be finite at the horizon.
This is equivalent to the requirement of finiteness of
\begin{equation}
F_{{\rm out}\,\hat{0}l}=\frac{F_{{\rm out}\,0l}}{N}=-\frac{1}{N}
\frac{
\partial \varphi_{\rm out}\,}{\partial l}\,,
\label{dervarphil}
\end{equation}
\begin{equation}
F_{{\rm out}\,\hat{0}a}=\frac{F_{{\rm out}\,0a}}{N}=-\frac{1}{N}
\frac{
\partial \varphi_{\rm out}\,}{\partial x^a}\,.
\end{equation}
But for a regular
horizon, the metric potential $N$ just outside the horizon must obey 
Eq.~(\ref{Nkappal-l01st}).
Therefore, from Eqs.~(\ref{Nkappal-l01st}) and~(\ref{dervarphil}),
we obtain that in this limit
\begin{equation}
\varphi_{\rm out} -\Phi=b(x^{a})(l-l_{\rm h})^{2}+
O((l-l_{\rm h})^{3})\, ,
\label{phi}
\end{equation}
for some function $b(x^{a})$,
where $\Phi\equiv\varphi_{\rm h} =\varphi_{\rm membrane\, at\, the\, horizon}
=\varphi_{\rm out\, at\, the\, horizon}
$
is constant, having the meaning of the potential at the membrane
when this is at the horizon.
By continuity, the
electric potential remains constant inside, in the vacuum region, with
value $\Phi$.
Thus, we have the important result that at the horizon the electric
potential is a constant
\begin{equation}
\Phi=\varphi_{\rm out\, at\, the\, horizon} ={\rm constant}\,.
\label{phiconstant}
\end{equation}

In addition, Eq.~(\ref{phi})
means that the normal component of $F_{\hat 0l}$ has a jump,
indeed, $\left( F_{{\rm out}\,\hat 0l}\right)_{+}$
is $\left( F_{{\rm out}\,\hat 0l}\right)_{+}=-\frac{1}{N}
\left(\frac{\partial \varphi_{\rm out}}{\partial l}
\right)_{+}=-\frac{2b}{\kappa}$,
and 
$\left( F_{{\rm in}\,\hat 0l}\right)_{-}=-\frac{1}{N}
\left( \frac{\partial \varphi_{\rm in}}
{\partial l}\right)_{-}=0$,
where again the subscripts
$+$ and $-$ mean the evaluation at the horizon from the outside and
the inside, respectively.
So, $\left( F_{{\rm out}\,\hat 0l}\right)_{+}-
\left( F_{{\rm in}\,\hat 0l}\right)_{-}=-\frac{2b}{\kappa}$,
i.e., there is a jump in the normal electric component.
Thus, comparing with the formulas above, see Eqs.~(\ref{gauss0})
and~(\ref{dervarphil}), i.e., $F_{{\rm ou}\,\hat 0l}=-4\pi\sigma_{\rm e}$,
we have $b={2\pi\sigma_{\rm e}}{\kappa}$ at the horizon.
On the other hand, the tangential components of the electric field
go as 
$F_{{\rm out}\,0a}\sim
(l-l_{\rm h})^{2}\rightarrow 0$,
and so
$F_{{\rm out}\,\hat{0}a}\sim (l-l_{\rm h})\rightarrow 0$.
Thus,  $\left( F_{{\rm out}\,\hat{0}a}\right)_{+}=0$ and
$\left( F_{{\rm in}\,\hat{0}a}\right)_{-}=0$.
We conclude that there is no jump
in the tangential electric components.

\subsubsection{Black hole mass formula in the membrane
paradigm with electric field}

Now, to find the black hole mass formula, we
start by using Eq.~(\ref{massmembranegeneral}).
Note that Eq.~(\ref{massmembranegeneral}) is independent of $N$.
Take then the limit 
$
N\rightarrow 0$,
and take into account the constancy of the potential in the horizon limit,
Eq.~(\ref{phiconstant}). Use Eq.~(\ref{totalcharge})
in Eq.~(\ref{massmembranegeneral})
to obtain
\begin{equation}
M_{\rm membrane \,at \,the \,horizon \,\,em }=\Phi\, Q\,,
\label{emhmass}
\end{equation}
where $M_{\rm membrane \,at \,the \,horizon \,\,em}$ is the
contribution from the electromagnetic mass of the membrane to the
black hole mass when this is at the gravitational radius.

In the outer region, if charged,
there is also a
contribution from the regular part of the current, namely,
$
M_{\rm out\,\,em}=\frac{1}{4\pi}\int_{\Sigma_{\rm out}}\varphi_{\rm out}\,
j_{\rm out}^{0}N\sqrt{\gamma}\, d^{3}x
$, see Eq.~(\ref{outerem1}), and where
$\Sigma_{\rm out}$ is the 3-space outside the horizon.
Further, the outer current can be written as
$j^{\mu}_{\rm out}
=\rho_{\rm out\,\,em}u^{\mu}$, and
$\rho_{\rm out\,\,em}u^{\mu}$ is an invariant charge
density, with $u^{\mu}$ being
the 4-velocity of the source.
In our case the only component of the current is $j^{0}_{\rm out}=
\frac{\rho_{\rm out\,\,em}}{N}$.
Then, 
\begin{equation}
M_{\rm em\,out}=\frac{1}{4\pi}\int_{\Sigma_{\rm out}}\varphi_{\rm out}
\,\,\rho_{\rm out\,\,em}\sqrt{\gamma}\, d^{3}x\,,
\label{massemout}
\end{equation}
valid outside the horizon.

Thus,
the total mass $M$ of a black hole is the sum of the
mass formula for a nonelectric black hole given in Eq.~(\ref{tot}),
plus the electric mass coming from membrane
at the horizon given in Eq.~(\ref{emhmass}),
plus the outer mass involving the sum of
Eq.~(\ref{out}) of all the matter and other nonelectric
fields outside the black hole
with the outer electric mass given by Eq.~(\ref{massemout}).
Summing all these mass contributions one finds the black hole mass
formula with an electric field, namely,
\begin{equation}
M=\frac{1}{4\pi}\kappa A+  \Phi\,Q+M_{\rm out}\,.
\label{totmassem}
\end{equation}
Equation~(\ref{totmassem}) is the mass formula for electric black holes,
now derived in the membrane paradigm.  This black hole mass formula
obtained through the membrane paradigm approach is the same as that
obtained by other methods \cite{bch} (see also \cite{cart73,bard73}).
When $M_{\rm out}=0$, Eq.~(\ref{totmassem})
is the Smarr formula for a static
and electric,
i.e.,  Reissner-Nordstr\"om,
black hole in general relativity \cite{sm}.

\subsubsection{Extremal electric case and the interpretation for the
surface gravity}

The extremal case corresponds to $\kappa =0$. Thus, 
\begin{equation}
M= \Phi\, Q+M_{\rm out}\,.
\label{totmassemextremal}
\end{equation}
For $M_{\rm out}=0$ and for Reissner-Nordstr\"om, one finds $\Phi=1$
and $M=Q$, which is the well-known relation for electrical extremal
black holes.  In principle, this derivation for the extremal case
through the limits suffices.  However, if one prefers, a more
consistent way of derivation in the extremal case requires taking into
account another asymptotic form of the lapse function for small
$N$. By definition, the extremal case implies that for a black hole
$N\sim \exp (-\alpha l)$, where $\alpha$ is some positive constant and
$l\rightarrow \infty $. Then, the above equations show that the
membrane stresses remain finite but their contribution to the surface
mass vanishes because of the form of the lapse function $N$. As a
result, indeed, $M_{\rm membrane\,at\,the\,horizon}=0$.

Since in the extremal case $\kappa =0$, our interpretation for
$\kappa$ implies that the horizon now has no surface energy density,
all the energy density it has comes from the electric field.

\section{Black hole mass formula in the membrane paradigm:
Rotating membrane}
\label{3}

\subsection{Preliminaries}

\subsubsection{Gravitational and rotational fields}

A  mass formula for a spacetime
with a rotational axially
symmetric background can also be derived
using the membrane paradigm. The considerations follow the same lines as before.
We
restrict ourselves to the electrically uncharged case.

A general rotating axially symmetric metric can be written
in $(t,r,\theta,\phi)$ coordinates as
\begin{equation}
ds^{2}=-N^{2}dt^{2}+g_{ll}dl^{2}+g_{\theta\theta}d\theta^{2}+
g_{\phi \phi}(d\phi -\omega 
dt)^{2}\,,
\label{gener}
\end{equation}
where the metric functions $N$,
$g_{ll}$,
$g_{\theta\theta}$,
$g_{\phi \phi}$, and 
$\omega$
are functions of $l$ and $\theta$ in general.
We assume there is a rotating shell that separates the vacuum, Minkowski,
inside spacetime from the outside spacetime, and 
thus from Eq.~(\ref{gener}),
we can specify the inside and outside metrics.

The metric for the vacuum, Minkowski, inner spacetime is written,
putting $g_{ll}=1$, as
\begin{equation}
ds^{2}=-N_0^{2}dt^{2}+dl^{2}+g_{\theta\theta}d\theta^{2}+
g_{\phi \phi}(d\phi -\omega_0 
dt)^{2} \,,
\label{metrotin}
\end{equation}
where $N_0$ and $\omega_0$ are conveniently chosen constants.
The metric for the spacetime outside the membrane is, putting also $g_{ll}=1$,
written as
\begin{equation}
ds^{2}=-N^{2}dt^{2}+dl^{2}+g_{\theta\theta}d\theta^{2}+
g_{\phi \phi}(d\phi -\omega 
dt)^{2}\,,
\label{metrotout}
\end{equation}
where $N$, $g_{\theta\theta}$, $g_{\phi \phi}$, and
$\omega$ are functions of $l$ and $\theta$
in general.

\subsubsection{Energy-momentum tensor}

As in the static case, we assume
that the spacetime has some energy-momentum tensor $T_{\mu\nu}$
and  divide the energy-momentum tensor as
the sum of inner energy-momentum tensor $T_{{\rm in}\,\mu\nu}$,
the membrane energy-momentum tensor $T_{{\rm membrane}\,\mu\nu}$,
and the outer energy-momentum tensor
$T_{{\rm out}\,\mu\nu}$,
\begin{equation}
T_{\mu\nu}=T_{{\rm in}\,\mu\nu}+T_{{\rm membrane}\,\mu\nu}+T_{{\rm out}\,\mu\nu}
\label{energmomenttotalrot}
\end{equation}
Inside is Minkowski, and so
inside,
\begin{equation}
T_{{\rm in}\,\mu\nu}=0\,.
\label{energmomentin1rot}
\end{equation}
The membrane is infinitesimal, situated at $l_0$,
so we put 
\begin{equation}
T_{{\rm membrane}\,\mu\nu}=S_{\mu\nu}\delta(l-l_0)\,,
\label{tmunumembrane1rot}
\end{equation}
where $\delta(l-l_0)$ is the Dirac delta function and $S_{\mu\nu}$
is a surface energy-momentum tensor defined at the membrane.
The outside energy-momentum tensor is divided
into matter and other fields, 
so
\begin{equation}
T_{{\rm out}\,\mu\nu}=T_{{\rm out\,\,matter}\,\mu\nu}+
T_{{\rm out\,\,other\,fields}\,\mu\nu}\,.
\label{energmomentouter1rot}
\end{equation}
The term $T_{{\rm out\,\,other\,fields}\,\mu\nu}$
can contain an electromagnetic field,
for instance.

\subsection{Tolman mass and angular momentum for a rotating membrane}

We start with the angular momentum as it will be useful for the
expression for the mass.
The Tolman spatial vector momentum is defined as $J_i=\int T
_{i}^{0}\,\sqrt{-g}\,d^{3}x$,
which in our case from the metric Eq.~(\ref{metrotout}) can be put in
the form  $J_i=\int T_{i}^{0}\,N\sqrt{g_3}\,d^{3}x$, where we have used
$-g=N^2g_3$, $g_3$ being the determinant of the spatial 3-metric.
The only momentum that matters here is the angular momentum $J_\phi$.
Moreover, $g_3=g_{\theta\theta}g_{\phi\phi}$ from
the metric Eq.~(\ref{metrotout}). So, defining
$\gamma=g_{\theta\theta}g_{\phi\phi}$,
we write 
the total angular momentum $J_\phi$ as given by
\begin{equation}
J_\phi=-\int T_{\phi}^{0}\,N\sqrt{\gamma}\,d^{3}x\,.
\label{Jstationary}
\end{equation}
Now, the total value of the angular momentum (\ref{Jstationary}) can be
split into three contributions, namely,
the inner vacuum, the membrane, and the outer region, such
that,
\begin{equation}
J_\phi=J_{\phi\,\,{\rm in}}+J_{\phi\,\,{\rm membrane}}+J_{\phi\,\,{\rm out}}\,.
\label{jtotal}
\end{equation}
For the inside, since
there is no matter inside, see Eq.~(\ref{energmomentin1rot}),
we have from Eq.~(\ref{Jstationary})
\begin{equation}
J_{\phi\,\,\rm in}=0\,.  \label{jint}
\end{equation}
For the membrane, its angular momentum can be calculated
through the expression, 
\begin{equation}
J_{\phi\,\,\rm membrane}=-\int_{\rm membrane}
T_{\phi}^{0}\,N\sqrt{\gamma}\,d^{3}x\,.
\label{Jmembranestationary1}
\end{equation}
Now, from Eq.~(\ref{tmunumembrane1rot}), we have
$T_{\rm membrane\,\phi}^{\quad\quad\quad\quad 0}=S_{\phi}^{\;0}\,\delta(l-l_0)$.
Since $8\pi
S_{\mu}^{\nu}=[[K_{\mu}^{\nu}]]-\delta_{\mu}^{\nu
}[[K]]$, we have here,
$8\pi
S_{\phi}^{0}=[[K_{\phi}^{0}]]-\delta_{\phi}^{0}[[K]]$, i.e.,
$8\pi
S_{\phi}^{0}=[[K_{\phi}^{0}]]$ and 
since the inside is flat, we have
$[[K_{\phi}^{0}]]=K_{+\,\phi}^{\quad 0}$.
We suppress the $+$ index in the following as it is
obvious that all quantities are calculated for the outside.
For the outside we have
$K_{\phi}^{0}=K_{\phi0}g^{00}+K_{\phi\phi}g^{0\phi}$.
Since $K_{\mu\nu}=-\nabla_\nu n_\mu$
and  $n_\mu=\frac{\partial l}{\partial x^\mu}$ we can calculate
$K_{\phi0}$ and $K_{\phi\phi}$.
We find
$K_{\phi0}=+\,\Gamma^l_{\phi0}n_l=
-\frac12 g^{ll}\frac{\partial g_{\phi0}}{\partial l}=
\frac12\left( g_{\phi\phi}
\frac{\partial \omega}{\partial l}+\omega
\frac{\partial  g_{\phi\phi}}{\partial l}\right)$,
where we have used $g^{ll}=1$, $n_l=1$, and 
$g_{\phi0}=-\omega g_{\phi\phi}$.
Also, 
$K_{\phi\phi}=+\Gamma^l_{\phi\phi}n_l=-\frac12 g^{ll}
\frac{\partial g_{\phi\phi}}{\partial l}
=-\frac12\frac{\partial g_{\phi\phi}}{\partial l}$.
In addition,
$g^{00}=-\frac{1}{N^2}$, and 
$g^{0\phi}=-\frac{\omega}{N^2}$. Then,
$K_{\phi}^{0}=-\frac12\frac{g_{\phi\phi}}{N^2}
\frac{\partial \omega}{\partial l}$,
and so $8\pi
S_{\phi}^{0}=-\frac12\frac{g_{\phi\phi}}{N^2}
\frac{\partial \omega}{\partial l}$.
All quantities are evaluated at the membrane $l=l_0$
from the outside.
Integrating through the delta function, we obtain
\begin{equation}
J_{\phi\,\,\rm membrane}=\int_{\rm membrane} j\,dS\,,
\label{Jmembranestationary2}
\end{equation}
where 
\begin{equation}
j\equiv\frac{1}{16\pi}
\frac{g_{\phi\phi}}{N}\frac{\partial \omega}{\partial l}\,,
\label{jmembranestationary22}
\end{equation}
is an angular momentum surface density, the quantities are evaluated
at the membrane from the outside,
and $dS$ is the 2-dimensional surface spanned by
$t={\rm constant}$ and
$l={\rm constant}$.
For the outside, when there is matter and other
fields outside the membrane one has
\begin{equation}
J_{\rm out}=-\int_{\rm out} T_{\phi}^{0}\,N\sqrt{\gamma}\,d^{3}x\,.
\label{Jstationaryout1}
\end{equation}

Now, we evaluate the mass in the stationary case,
resorting  to the Tolman mass formula.
The Tolman mass for a given  stationary spacetime
with a stress-energy tensor $T_{\mu\nu}$ 
is defined as in the static case, see Eq.~(\ref{mass4}), i.e.,
\begin{equation}
M=\int_\Sigma(-T_{0}^{0}+T_{k}^{k})\sqrt{-g}d^{3}x\,,
\label{mass4stationary}
\end{equation}
with $g$ being the determinant of the metric $g_{\mu\nu}$
in, e.g., Eq.~(\ref{metrotout}),
$d^3x=dx^1dx^2dx^3$,
and the integral is performed
over a 3-space $\Sigma$ with $t={\rm constant}$.
Using, from
the metric Eq.~(\ref{metrotout}), that $g=-N\gamma$ 
with $\gamma=g_{\theta\theta}g_{\phi\phi}$,
we have
\begin{equation}
M=\int_\Sigma(-T_{0}^{0}+T_{k}^{k})\,N\,\sqrt{\gamma}d^{3}x\,.
\label{massgammastationary}
\end{equation}
Again, we divide the mass into three pieces,
\begin{equation}
M=M_{{\rm in}}+M_{{\rm membrane}}+M_{{\rm out}}\,.
\label{mtotalstatio}
\end{equation}
For the inside vacuum, we have
\begin{equation}
M_{\rm in}=0\,.
\label{mintstationary}
\end{equation}
For the membrane, we
apply the junction conditions. 
Since $8\pi
S_{\mu}^{\nu}=[[K_{\mu}^{\nu}]]-\delta_{\mu}^{\nu
}[[K]]$, we have here $8\pi
\left(-S_{0}^{0}+S_{a}^{a}\right)=-2[[K_{0}^{0}]]$.
Put $n^{\mu}$ as the unit
vector normal to the boundary surface, the membrane.
Also, $K_{\mu \nu}=-\nabla_\nu n_\mu$, where at the
boundary surface $N={\rm constant}$ and the normal unit vector is $
n_{\mu}=\frac{\partial l}{\partial x^\mu}$.
Further
calculations then give
\begin{equation}
M_{\rm membrane}=\int_{\rm membrane}
\left(
\sigma+2\,\omega\, j
\right) \,dS \,,
\label{alstationary}
\end{equation}
where $\sigma$ has the expression given in  Eq.~(\ref{muresult}),
$\omega$ is the membrane angular velocity
with $\omega=\omega_0$ defined in 
Eq.~(\ref{metrotin}), and
$j$ is given in Eq.~(\ref{jmembranestationary22}).

For the outside, we have
\begin{equation}
M_{{\rm out}}=\int_{\rm out}\left(-{T_{\rm out\,0}}^{\,0}+
{T_{\rm out\,k}}^{\,k}\right)
\,N\,\sqrt{\gamma}\,d^{3}x\,,
\label{outrot}
\end{equation}
where one can further split into matter fields and other fields.

Thus, putting together
Eqs.~(\ref{mintstationary}), (\ref{alstationary}), and~(\ref{outrot}) into 
Eqs.~(\ref{mtotalstatio}),
we obtain the mass formula from the Tolman definition for a
spacetime with a membrane.

\subsection{Black hole mass formula in the membrane paradigm with rotation}

\subsubsection{Constancy of the angular velocity at the horizon}

Now, let $l_{\rm h}\equiv l_{0\,\rm at\, the\, horizon \,limit}$.
The membrane is approaching the horizon.
In the nonextremal case we have
that $N$ behaves as given in Eq.~(\ref{Nkappal-l01st}), see \cite{visprd}.
It still involves $\kappa$, the surface gravity at the horizon obeying $\kappa =
{\rm constant}$ different from zero.
For the metric coefficient $\omega$, we assume the validity of the expansion
near the horizon \cite{visprd,tz}
$
\omega =\Omega+a_{2}N^{2}+...\,,
$
for some constant $a_2$, and so from Eq.~(\ref{Nkappal-l01st}),
we get
\begin{equation}
\omega =\Omega+O((l-l_{\rm h})^{2})\,, \label{om}
\end{equation}
where $\Omega\equiv \omega_{\rm h}$ is the horizon value of
$\omega $ obeying
\begin{equation}
\Omega={\rm constant}\,.  \label{w2}
\end{equation}

\subsubsection{Black hole mass formula in the membrane
paradigm with rotation}

To obtain the black hole mass formula in the membrane paradigm 
with rotation,
let us focus on the membrane term above. Using Eq.~(\ref{alstationary})
and with the help of 
Eqs.~(\ref{muresult}) and~(\ref{jmembranestationary22}),
we can put the $\sigma$ term, now in the guise of $\kappa$, and the
$\omega$ term, now as $\Omega$, outside the integral,
since $\kappa$ and $\Omega$ are  constants
at the horizon.
For the $\kappa$ term we get $\frac{\kappa A}{4\pi}$
after integration,
with $A$ being the horizon area.
For the $\Omega$ term we get $2\,\Omega\,J$
after integration, where 
$J$ is the the angular momentum of the membrane at the horizon,
i.e., the angular momentum of the black hole.
So,  Eq.~(\ref{alstationary}) at the horizon
together with Eq.~(\ref{outrot})
and  Eq.~(\ref{mintstationary})
in Eq.~(\ref{mtotalstatio}),
lead to 
\begin{equation}
M=\frac{\kappa A}{4\pi}+2\,\Omega\,J+M_{\rm out}\,.
\label{mbhrot}
\end{equation}
This is precisely the black hole mass formula for rotating
black holes derived from the
membrane paradigm.
It was derived in \cite{bch} (see also \cite{cart73,bard73})
from black hole
theory using 
Komar's mass and 
angular momentum 
definitions, which are appropriate in a vacuum spacetime
with one timelike and one spacelike
Killing vector.
In the vacuum case, $M_{\rm out}=0$, Eq.~(\ref{mbhrot})
is the Smarr formula for a Kerr
black hole \cite{sm}.

\subsubsection{Extremal rotating case}

For the extremal rotating case, one has $N=O((l-l_{\rm h})^3)$
\cite{visprd}.   For the metric coefficient $\omega$, we
assume the validity of the expansion near the horizon \cite{visprd,tz} $
\omega =\Omega+a_{1}N+...  $, i.e., $ \omega =\Omega
+O((l-l_{\rm h})^3)+...  $.  This all leads to $\kappa=0$, so that we
can use Eq.~(\ref{mbhrot}) directly and write the mass formula for the
extremal case as
\begin{equation}
M=2\,\Omega\,J+M_{\;{\rm out}}\,.
\end{equation}
For $M_{\;{\rm out}}=0$ and for an extremal Kerr, one finds $\Omega=\frac{1}{2M}$
and $M^2=J$, 
which is the well-known relation for rotating extremal black holes.

\subsubsection{Black hole mass formula in the membrane
paradigm with electric field and rotation}

If the membrane has electric field and rotation, developing the
calculations done previously, one finds the following
mass formula, 
\begin{equation}
M=\frac{\kappa A}{4\pi}+\Phi\, Q+2\,\Omega\, J
+M_{\;{\rm out}}\,.
\end{equation}
In the vacuum case, $M_{\;{\rm out}}=0$,
it is the Smarr formula for a Kerr-Newman
black hole \cite{sm}.

\section{Conclusions}
\label{conc}

In replacing the black hole event horizon by a self-gravitating
material membrane located slightly above the horizon itself, i.e.,
using the membrane paradigm formalism, we have been able to derive the
mass formula for static and rotating black holes without and with
electric fields.  A fundamental element in the derivation is the
setting of the proper boundary conditions on the membrane when it
approaches its own gravitational or horizon radius.  We found that
both the membrane paradigm and the standard black hole theory yield
exactly the same result.

We have thus filled an important gap in the
membrane paradigm investigations and showed that the membrane
paradigm formalism is able to
reproduce correctly one of the basics of black hole physics, namely,
the mass formula. This  has been achieved in a completely
different perspective than in the original derivations since now there
is no a black hole and we have dealt with a timelike surface only in
the form of a membrane.

\begin{acknowledgments}

We thank Funda\c c\~ao para a Ci\^encia e Tecnologia (FCT), Portugal,
for financial support through Grant~No.~UID/FIS/00099/2013.~JPSL
thanks FCT for the Grant No.~SFRH/BSAB/128455/2017 and Coordena\c
c\~ao de Aperfei\c coamento do Pessoal de N\'\i vel Superior (CAPES),
Brazil, for support within the Programa CSF-PVE, Grant
No.~88887.068694/2014-00.  OBZ thanks support from SFFR, Ukraine,
Project No.~32367 and thanks also Kazan Federal University for a state
grant for scientific activities.

\end{acknowledgments}

\end{document}